\documentclass[sigconf,screen]{acmart}
\usepackage{csquotes}
\usepackage{tablefootnote}
\usepackage{booktabs}
\AtBeginDocument{%
  }

\begin{document}

\title{Leveraging LLMs for User Stories in AI Systems: UStAI Dataset}

\author{Asma Yamani}
\email{g201906630@kfupm.edu.sa}
\orcid{0000-0002-6277-8972}
\affiliation{%
  \institution{King Fahd University of Petroleum and Minerals}
  \city{Dhahran}
  \country{KSA}
}

\author{Malak Baslyman}
\email{malak.baslyman@kfupm.edu.sa}
\orcid{0000-0003-4002-4480}
\affiliation{%
  \institution{King Fahd University of Petroleum and Minerals}
  \city{Dhahran}
  \country{KSA}
}

\author{Moataz Ahmed}
\email{moataz@kfupm.edu.sa}
\orcid{0000-0003-0042-8819}
\affiliation{%
  \institution{King Fahd University of Petroleum and Minerals}
  \city{Dhahran}
  \country{KSA}
}

\renewcommand{\shortauthors}{Yamani et al.}

\begin{abstract}
AI systems are gaining widespread adoption across various sectors and domains. Creating high-quality AI system requirements is crucial for aligning the AI system with business goals and consumer values and for social responsibility. However, with the uncertain nature of AI systems and the heavy reliance on sensitive data, more research is needed to address the elicitation and analysis of AI systems requirements. With the proprietary nature of many AI systems, there is a lack of open-source requirements artifacts and technical requirements documents for AI systems, limiting broader research and investigation. With Large Language Models (LLMs) emerging as a promising alternative to human-generated text, this paper investigates the potential use of LLMs to generate user stories for AI systems based on abstracts from scholarly papers.  We conducted an empirical evaluation using three LLMs and generated $1260$ user stories from $42$ abstracts from $26$ domains. We assess their quality using the Quality User Story (QUS) framework. Moreover, we identify relevant non-functional requirements (NFRs) and ethical principles. Our analysis demonstrates that the investigated LLMs can generate user stories inspired by the needs of various stakeholders, offering a promising approach for generating user stories for research purposes and for aiding in the early requirements elicitation phase of AI systems. We have compiled and curated a collection of stories generated by various LLMs into a dataset (UStAI), which is now publicly available for use. 
\end{abstract}

\ccsdesc[500]{Computing methodologies~Artificial intelligence}
\ccsdesc[500]{Software and its engineering~Requirements analysis}

\keywords{User stories, large language models, quality requirements, requirements elicitation, requirements generation}



\maketitle
\section{Introduction}
Artificial intelligence (AI) systems are ultimately software systems with non-deterministic components~\cite{10.1109/MS.2020.2985621}. This leads to unique challenges when applying conventional software practices, especially concerning requirements engineering (RE), testing, and addressing ethical issues~\cite{Halme2024,10.1145/3487043}. With the proprietary nature of many AI systems, there is a lack of open-sourced requirements artifacts concerning AI-based systems~\cite{10.1145/3597503.3639185}, limiting the ability of broader academic and industrial communities to learn from, investigate, and build upon them. 

Large Language Models (LLMs) have emerged as a promising alternative or supplement to human-generated text to circumvent these limitations~\cite{long-etal-2024-llms}. They can mimic the characteristics and patterns of real-world data and leverage the vast repository of knowledge acquired in their pre-training step. LLMs have been introduced in multiple domains suffering from data scarcity, such as medical data~\cite{kumichev2024medsynllmbasedsyntheticmedical}. LLMs were also used to generate requirements in the form of user stories. Ferrara et al.~\cite{10.1145/3597503.3639185} used LLM to construct an AI user stories dataset based on an ontology of domains and a set of ML algorithms. Abed et al.~\cite{10.1007/978-3-031-62110-9_1} generated user stories based on stakeholders' interviews.
There is a lack of public RE datasets containing requirements relevant to AI-based systems. The current dataset available~\cite{10.1145/3597503.3639185} is a highly synthetic dataset and is not annotated with common RE tasks (e.g, non-functional requirements (NFRs) detection, ambiguity detection, conflict detection etc). Studies investigating the use of LLM in eliciting or generating requirements focused on a limited number of cases or systems, and did not result in public datasets~\cite{bano2023aialloperationalisingdiversity,10.1007/978-3-031-62110-9_1,10795004,10724709,10795004,10109345}.\par
Scholarly abstracts are unique in providing a large accessible source for descriptions on the ML component of AI-based systems. Abstracts capture high-level problem statements that can help inform user needs and scenarios. They also provide assumptions and methodology to help set realistic expectations for the goal. Moreover, research outcomes are, in many cases, proof-of-concepts to what can evolve to a complete ML systems. However, their feasibility to be an input to the user story generation process has not been investigated. In search of diverse and realistic high-quality AI user story corpora, we seek to answer the following research questions in this work.
\begin{enumerate}
    \item  Can Large Language Models be used to generate user stories from the perspective of multiple stakeholders based on a scholarly abstract description of the AI component?
    \item Which LLM generates the highest quality user stories according to the Quality User Story (QUS) framework~\cite{lucassen2016improving}?
    \item To what extent do LLM-generated user stories capture ethical considerations and principles and non-functional requirements (NFRs) relevant to AI systems?

\end{enumerate}
We summarize our contributions as follows. \begin{itemize} \item Propose the use of LLMs to generate user stories based on abstracts of scholarly papers describing AI components. \item Perform an empirical evaluation of $1260$ generated user stories. \item Compile and curate a collection of user stories for AI systems dataset (UStAI)\footnote{https://github.com/asmayamani/UStAI} with $3000$ user stories including $1260$ user stories evaluated for quality attributes from the QUS framework and non-functional requirements, including implied ethical principles. \end{itemize}

The remainder of this paper is structured as follows. Section~\ref{rw} presents related work, discussing prior research on using Large Language Models (LLMs) for requirements generation and available datasets for requirements engineering. Section~\ref{sd} describes the study design, including the methodology for dataset generation, selection of abstracts, and evaluation framework. Section~\ref{r} presents the results of our empirical evaluation, including the quality assessment of generated user stories and their alignment with ethical principles and non-functional requirements. Section~\ref{d} discusses the findings, comparing the quality of user stories across different LLMs and highlighting potential dataset applications. Section~\ref{c} concludes the paper with a summary of key insights and contributions.

\section{Related Work}
\label{rw}
In this section, we will present related work on the use of LLMs in requirements generation. Also, enumerate datasets curated for requirements engineering research.
\subsection{LLMs in requirements generation}
AI-based techniques and tools have been investigated to aid in the generation of user stories. Rodeghero et al.\cite{7985649} used Recurrent Neural Networks (RNNs) to identify roles, features, and motivations from customer interviews. The approach achieved 70.8\% precision and 18.3\% recall. Dwitma and Rusli\cite{Dwitama2020} used chatbots that utilize Artificial Intelligence Markup Language to elicit specific information from various stakeholders. This approach was used to generate user stories. However, since user stories are context-specific, this approach is limited and requires extensive tuning.

Utilizing the embedded knowledge in the pretraining step in LLMs, Sharma et al.~\cite{10724709} investigated the ability of LLMs, GPT-3.5-turbo in particular, to generate user stories based on problem description using a 4-step chain prompt. They evaluated their approach on four problem descriptions. GPT-3.5-turbo, using their approach, achieved a recall of 96\% and 81\% for the role and function parts, respectively. Abed et al.~\cite{10.1007/978-3-031-62110-9_1} evaluated the use of LLMs to generate user stories from a backlog of interviews with stakeholders, using a subset of the QUS framework~\cite{lucassen2016improving}. The highest recall was 61\%, achieved by PRD Maker and GPT4, covering 11 features out of the 18 identified by the company.

Rahman et al.~\cite{10795004} proposed a 3-step prompt approach to extract user stories and appropriate test cases from requirements specification documents. In the first step, the requirements document is passed through the prompts into the LLM to extract the functional and non-functional requirements. In the second shot, the LLM generates the deliverables with a clear definition of task specifications, technical details, and acceptance criteria. Lastly, the third shot the LLM to generate the test coverage specifications. They used GPT-4.0-turbo to evaluate their approach on six mid-size RE documents from different domains. The majority of 76 developers evaluated the generated user stories as good (4 out of 5) on a survey to evaluate the readability, understandability, specificity, and technical aspects. More improvement is required, specifically in the Specificity and Technical Aspects categories. They released a tool called Genius for public use.

\emph{While LLMs were investigated for the elicitation and generation of requirements from multiple sources ~\cite{10795004,10724709,10795004,10109345}, the use of scholarly abstracts as an input to the generation was not investigated. Abstracts capture high-level problem statements and innovative approaches that imply underlying user needs. They detail the ML methodologies and constraints that are critical for system implementation. In addition, it provides an early insight into how cutting-edge research can meet user needs. This input source is specifically valuable for AI-based systems as many AI systems have roots in research conducted by data scientists, and many AI-based systems begin as proof-of-concept models developed to test hypotheses.}

\begin{table*}[]
\caption{A summery of the available datasets for requirements artifacts.\label{da}}
\resizebox{\linewidth}{!}{%
\begin{tabular}{p{0.08\linewidth}p{0.19\linewidth}p{0.2\linewidth}p{0.08\linewidth}p{0.07\linewidth}p{0.2\linewidth}p{0.05\linewidth}p{0.29\linewidth}}
\toprule
\textbf{Dataset title {[}ref{]}}                                                                                                                                                                                            & \textbf{Source}                                                                                & \textbf{\#records}                                                                                                          & \textbf{Format(or structure)}                          & \textbf{Contains ML Projects?} & \textbf{Annotation}                                                                                               & \textbf{Data source Year(s)} & \textbf{Tasks}                                                                                                                                                                                                                                                                 \\ \midrule
\textbf{PROMISE NFR\cite{boetticher2007promise}}                                                                                                 & Collected from publicly available Software Requirement Specification (SRS) documents.          & 625 labeled software requirements (443 FR and 526 NFRs)                                                                     & Tabular data                                           & No                             & FR/NFR                                                                                                            & 2012\textgreater{}         & Requirements Classification. It can be further annotated to work as a benchmark for other tasks, such as ambiguity detection, requirements categorization and identification of equivalent requirements.                                                                       \\
\textbf{PROMISE-EXP~\cite{lima2019software}}                                                                                                                                  & Expands on PROMISE NFR collected from publicly available Software Requirement Specification (SRS) documents.          & 969 labeled software requirements (443 FR and 526 NFRs)                                                                     & Tabular data                                           & No                             & FR/NFR                                                                                                            & 2019\textgreater{}         & Requirements Classification. It can be further annotated to work as a benchmark for other tasks.                                                                       \\

\textbf{PURE~\cite{Ferrari2017PURE}}                                                                                                                              & Collected from publicly available Software Requirement Specification (SRS) documents.          & 79 publicly available natural language requirements documents collected from the Web. The dataset includes 34,268 sentences & 17 in XML documents; the rest are in a document format & No                             & -                                                                                                                 & 1999-2011                  & It can be used for natural language processing tasks that are typical in requirements engineering. It can be further annotated to work as a benchmark for other tasks. \\
\textbf{ReqList~\cite{Sahu2024ReqNet}}                                     & PURE dataset (after applying inclusion/exclusion criteria) + more publicly available dataset & 86 systems, with a total of 12701 requirements                                                                              & Plain text                                             & No                             & The requirements are annotated implicitly by retaining their clause numbering and hierarchical structure          & 1995-2022                  & It can be used for natural language processing tasks that are typical in requirements engineering. It can be further annotated to work as a benchmark for other tasks. \\
\textbf{ReqNet~\cite{Sahu2024ReqNet}}                                                                                                                                          & Constructed from ReqList                                                                       & 17375 nodes                                                                                                                 & Graph                                                & No                             & The graph is annotated automatically by the structure derived from the clause numbers and document hierarchy      & 1995-2022                  & Graph-theoretic analyses (e.g., traceability, change propagation, and impact analysis)                                                                                                                                                                                         \\
\textbf{ReqSim~\cite{Sahu2024ReqNet}}                                                                                                                                      & Constructed from ReqList                                                                       & 10933 pairs of requirements with similarity scores                                                                                                 & Tabular data                                           & No                             & The similarity scores are computed semi-automatically using weighted distances from the extracted tree structures & 1995-2022                  & sentence-level semantic similarity analysis                                                                                                                                                                                                                                    \\

\textbf{RETRO~\cite{Huffman2018REquirements}}                                                                                                            & Industry                                                                                       & 66 high level requirements for the requirement specification of RETRO tool and an answer set containing a total of 301 links for tracing the artifacts to each other                                              & Compatible with RETRO.NET traceability software        & No                             & Traceability matrix (TM), a collection of trace links                                                             & 2004-2006                  & Traceability-related tasks such as trace link generation, trace matrix assessment, and satisfaction assessment.                                                                                                                                                                \\
\textbf{\cite{Nazim2020Generating}}                             & Case study done for an Institute Examination Systems from academia                                                                                       & 32 functional requirements for an Institute Examination Systems, each annotated with importance of 3 NFR                                                         & The dataset is in the paper                            & No                             & Importance (high, medium, low) of 3 NFRs (Usability, Cost, and Security)                                          & 2020                       & Requirements prioritization                                                                                                                                                                                                                                                     \\
\textbf{\cite{Dalpiaz2018}}                                                                                                                                 & Collected from publicly available dataset                                                    & 22 user story sets. Each set include 50+ requirements, with a total of 2,067 user stories                                   & Plain text                                             & No                             & -                                                                                                                 & 2004-2018                  & Originally used to conduct experiments about ambiguity detection with the REVV-Light tool. It can be further annotated to work as a benchmark for other tasks.        \\
\textbf{SecReq~\cite{Knauss2021}}                                                                                                                                  & Collected from publicly available dataset                                                    & 3 requirements sets. Each set include 100+ requirements, with a total of 510 requirements                                   & Tabular data                                           & No                             & Security relevant requirement                                                                                     & 2007-2010                  & Used initially to identify security-relevant requirements. It can be further annotated to work as a benchmark for other tasks.                                                                                                                                                  \\
\textbf{ReFair~\cite{10.1145/3597503.3639185}}                                                                      & Synthetic using ChatGPT-4 based on an ontology                                                 & 12,401 user stories across 34 application domains                                                                          & Tabular data                                           & Yes                            & Domain, ML Task, context-specific sensitive features                                                              & 2024                       & Domain Classification, ML Task classification, recommend context-specific sensitive features                                                                                                                                                                                   \\
\textbf{UStAI  (Ours)}                                                                                                                              & Synthetic using 3 LLMs: ChatGPT o1-mini, Gemini 1.5 flash, Llama 3.1 70b                       & 3000 user stories from 100 ML systems, including 1260 evaluated for quality attributes and annotated with NFRs and ethics requirements                                                                                               & Tabular data                                           & Yes                            & Quality according to the QUS framework, High-level NFR, indicated ethics principles.                              & 2024-2025                  & Quality evaluation, ethical issues identification, ambiguity detection, requirements categorization, conflict detection, and identification of equivalent requirements.                                                                                                      \\ \bottomrule
\end{tabular}
}
\end{table*}

\subsection{Datasets for requirements engineering}
The availability of high-quality requirements datasets is crucial in advancing the automation and development of predictive models and solutions in requirements engineering. This includes training NLP, graph, and machine learning models to tackle ambiguity detection, traceability, conflict resolution, and many other tasks~\cite{Sahu2024ReqNet,Ferrari2017PURE}. Despite the advancement of Generative AI (GenAI), requirements datasets are still needed to evaluate GenAI solutions. Curating comprehensive and well-annotated datasets has been a challenge in RE~\cite{Sahu2024ReqNet}. A summary of some of the available datasets for RE is in Table~\ref{da}.
The PROMISE repository was one of the first large-scale efforts to publicly share software engineering datasets for empirical research~\cite{boetticher2007promise}. Developed in the mid-2000s, it provided foundational data for building and validating predictive models. Other datasets did spin from including more requirements or annotation~\cite{lima2019software}. PURE, published in 2017, is one of the pioneering collections of publicly available requirements documents curated specifically for natural language processing research in requirements engineering~\cite{Ferrari2017PURE}. ReqList~\cite{Sahu2024ReqNet} extracted and preprocessed the requirements from selected documents from PURE and 32 other requirements documents. It provided the requirements in plain text, preserving clause numbers and hierarchical metadata. ReqNet provided a graph representation of ReqList constructed by converting each document’s requirements into a tree. Moreover, ReqSim provided similarity scores calculated between ReqList requirement pairs~\cite{Sahu2024ReqNet}.\par
For task-specific datasets, SecRec provides a dataset annotated with security-relevant requirements~\cite{Knauss2021}. RETRO, a requirements traceability software, open-sourced its requirements along with traceability relevant annotation for traceability tasks~\cite{Huffman2018REquirements}. A collection of 22 user story sets was released as a part of a study on ambiguity detection~\cite{Dalpiaz2018}. For NFR prioritization, a dataset was released based on a case study done for an Institute Examination System containing the functional requirements for this system and the prioritization of three functional requirements for each functional requirement~\cite{Nazim2020Generating}.\par 

Although ML models have been integrated into many systems since early 2010, requirements datasets containing ML projects remain scarce. The only dataset that focuses on AI-based systems is work by Ferrara et al.~\cite{10.1145/3597503.3639185}. It used ChatGPT to synthetically generate AI-related user stories from an ontology that mapped domains and ML tasks. This work covered diverse contexts and resulted in a large dataset. However, each topic was covered from the perspective of a single stakeholder, and as the domains and ML tasks were combined from an ontology and fed into the prompt template, the user stories appeared highly synthetic and strictly followed the template in the prompt, limiting the diversity. It was also not evaluated in terms of quality and lacked non-functional and ethics-related requirements, which are crucial in the elicitation process.

\emph{Due to the lack of well-annotated requirements dataset containing AI-based systems, building on previous studies, this work investigates the feasibility of using LLMs to generate high-quality user story datasets from the perspective of multiple stakeholders for AI systems, including both functional and non-functional requirements from scholarly abstracts. 
}

\section{Study Design}
\label{sd}
\begin{table*}[]
\centering
\caption{User Stories Quality Assessment criteria \cite{lucassen2016improving}}
\label{tab:my-table}
\resizebox{\linewidth}{!}{%
\begin{tabular}{p{0.3\linewidth}p{0.8\linewidth}p{0.1\linewidth}}
\toprule
Criteria                                                                     & Description                                                                                     &       Criteria type     \\ \midrule
\textbf{Syntactic}                                                                                                                            \\
Well-formed                                                                  & A user story includes at least a role and a means                                               & Individual \\
Atomic                                                                       & A user story expresses a requirement for exactly one feature                                    & Individual \\
Minimal                                                                      & A user story contains nothing more than role, means, and ends                                   & Individual \\
\textbf{Semantic}                                                                                                               \\
Conceptually sound                                                           & The means express a feature, and the ends express a rationale                                   & Individual \\
Problem-oriented & A user story only specifies the problem, not the solution to it                                 & Individual \\
Unambiguous                                                                  & A user story avoids terms or abstractions that lead to multiple interpretations                 & Individual \\
Conflict-free                                                                & A user story should not be inconsistent with any other user story                               & Set        \\
\textbf{Pragmatic}                                                                   &                                                                                                 &            \\
Full-sentence                                                                & A user story is a well-formed full sentence                                                     & Individual \\
Estimatable                                                                  & A story does not denote a coarse-grained requirement that is difficult to plan   & Individual \\
Complete                                                                     & Implementing a set of user stories creates a feature-complete application & Set        \\
Unique                                                                       & Every user story is unique, duplicates are avoided                                              & Set        \\
Independent                                                                  & The user story is self-contained and has no inherent dependencies              &    Set        \\
Uniform                                                                      & All user stories in a specification employ the same template                                    & Set        \\ \bottomrule
\end{tabular}%
}
\end{table*}

\subsection{Requirements artifact type selection.} User stories are a type of requirements artifact that describes software system features from the perspective of the users interacting with the system being developed~\cite{10.1016/j.chb.2014.10.046}. Popularized by Mike Cohn~\cite{Cohn2004-zk}, user stories are typically written in the format “As a ⟨type of user⟩, I want ⟨goal⟩, so that ⟨some reason⟩,” known as the Connextra notation.  User stories were chosen as the relation between the user, feature, reason, or end is essential to study the requirements for a human-centric approach. 

\begin{table*}[htbp]
  \scriptsize
  \setlength{\tabcolsep}{2pt}
  \centering
  \caption{Annotated user stories for abstract with title Machine Learning-Based Aggression Detection in Children with ADHD Using Sensor-Based Physical Activity Monitoring~\cite{s23104949}(excerpt). \small{Abbreviation key: Cjn = conjunctions, Min = minimal, CS = conceptually sound, PO = problem‑oriented, UA = unambiguous, CF = conflict‑free, Est = estimatable, Ind = independent, Uniq = unique, dup = \texttt{isSemDuplicate}, n.v.r = \texttt{no\_valid\_role}, SO = solution‑oriented, Gen = general.}}
  \label{tab:a2-userstories}
  \resizebox{\textwidth}{!}{%
\begin{tabular}{p{0.05\linewidth}p{0.02\linewidth}p{0.05\linewidth}p{0.04\linewidth}p{0.28\linewidth}p{0.09\linewidth}p{0.1\linewidth}p{0.03\linewidth}p{0.03\linewidth}p{0.03\linewidth}p{0.03\linewidth}p{0.03\linewidth}p{0.03\linewidth}p{0.03\linewidth}p{0.02\linewidth}p{0.02\linewidth}p{0.03\linewidth}p{0.03\linewidth}p{0.02\linewidth}p{0.03\linewidth}}
    \toprule
    ID & US & LLM & Role &
    \multicolumn{1}{c}{User story} &
    \multicolumn{1}{c}{NFR} & \multicolumn{1}{c}{Ethics} &
    At. & Mi. & CS & PO & UA &
    CF & CF\textsubscript{us} &
    Est. & Indep. & Uniq. & Uniq.\textsubscript{us} &
    M/E & M/E\textsubscript{us} \\ \midrule

    A2US1Ge & US1 & Gemini & parent &
    As a parent, I want a device that can track my child's physical activity and potentially identify aggressive episodes so that I can intervene and provide support. &
    N/A & N/A &
    Cjn & Min & CS & PO & UA &
    CF & -- & Est & Ind & -- & -- & End & 8 \\

    A2US2Ge & US2 & Gemini & teacher &
    As a teacher, I want a tool that can discreetly monitor student activity levels in the classroom and alert me to potential aggression so that I can de‑escalate situations before they escalate. &
    N/A & N/A &
    Cjn & Min & CS & PO & UA &
    Conf & 9 & Est & No & -- & -- & End & 8 \\

    A2US9Ge & US9 & Gemini & advocate &
    As a data‐privacy advocate, I want assurances that any wearable device used for aggression detection collects data securely and protects children's privacy. &
    Privacy–Security & Privacy / NM / Sec &
    At & Min & CS & PO & Vag &
    Src & 2,3,4,5,8 & Est & Ind & Uniq & -- & -- & -- \\

    A2US1Ll & US1 & Llama & researcher &
    As a researcher, I want to develop an objective method to track physical‑aggressive incidents in children, so that I can identify early warning signs and provide timely interventions. &
    N/A & Beneficence &
    At & Min & n.v.r & SO & Gen &
    CF & -- & No & No & -- & -- & Mean & 10 \\

    A2US2Ll & US2 & Llama & parent &
    As a parent, I want to know if my child is experiencing aggressive episodes, so that I can provide support and guidance to help them manage their emotions and behaviors. &
    N/A & N/A &
    At & Min & CS & PO & UA &
    CF & -- & Est & No & dup & 7 & End & -- \\

    A2US7Ll & US7 & Llama & parent &
    As a parent, I want to receive alerts and notifications when my child experiences a physical‐aggressive incident, so that I can provide immediate support and guidance. &
    Reliability & N/A &
    At & Min & CS & PO & UA &
    CF & -- & Est & No & dup & 2 & End & 2 \\

    A2US1o1 & US1 & o1‑mini & parent &
    As a parent, I want to receive real‑time alerts when my child shows aggressive behavior, so I can intervene immediately. &
    Perf.–Reliab. & N/A &
    At & Min & CS & PO & UA &
    CF & -- & Est & No & -- & -- & End & -- \\

    A2US2o1 & US2 & o1‑mini & teacher &
    As a teacher, I want daily reports on students' aggression levels from wearable sensors, so I can implement effective classroom strategies. &
    N/A & N/A &
    At & Min & CS & PO & UA &
    Conf & -- & Est & Ind & diff & -- & End & -- \\

    A2US9o1 & US9 & o1‑mini & officer &
    As a privacy officer, I want all sensor data to be securely stored and access‑controlled, so children's privacy is protected. &
    Security–Privacy & Privacy / Sec &
    At & Min & CS & PO & UA &
    Src & -- & Est & Ind & Uniq & -- & -- & -- \\

    \bottomrule
  \end{tabular}}
\end{table*}

\subsection{Dataset generation.} 
\textbf{Abstract selection} Initially, we generated a total of $3000$ user stories from $100$ scholarly abstracts covering diverse domains. However, to ensure a detailed, rigorous, and feasible evaluation, we selected randomly a representative subset consisting of $1260$ user stories from $42$ abstracts for in-depth analysis and annotation in this study covering $26$ domains. $24$ of these domains were from the ontology developed by Fabris et al.~\cite{Fabris2022}, which includes topics related to information and computer science, health, social science, economics and business, natural science, and other domains. In addition, we included abstracts related to autonomous vehicles and security, which are not a part of the Fabris et al.~\cite{Fabris2022} ontology, to ensure wider coverage. Twenty-six abstracts were from conference papers, while the remaining 14 were from Q1 or Q2 journals. The abstracts represent a system summary, as the construct of the abstract has the content needed in the user story in terms of background to identify the stakeholders <user>, the methodology to identify the <goal>, and the conclusion to identify the <reason>. Abstracts also are available, accessible, and free. Moreover, the feasibility of many ML systems is explored through the development of ML components by ML researchers in academic institutes.\par

\textbf{LLM selection} Three LLMs are chosen: Gemini 1.5 Flash ~\cite{geminiteam2024geminifamilyhighlycapable}, ChatGPT using o1-mini~\cite{openai2024gpt4technicalreport}, and Llama 3.1 70b~\cite{touvron2023llamaopenefficientfoundation}. This selection of LLMs ensures diversity in LLM Families, variety in model Sizes and complexity, diversity of training sources, and different methods in safeguards and alignment. By combining the results of the LLM generation, we maximize the diversity in the generation of requirements, capturing a wide range of viewpoints.\par
\textbf{Prompt crafting} After experimenting with multiple prompts, the prompt structure used was: "As a requirements engineer, write ten user stories for the following abstract: <abstract>." By using this prompt, we first direct the LLM to assume the persona of an expert in requirements, providing more relevant output. Specifying the number of generated user stories to be ten is to incentivize the LLM to generate more user stories instead of the observed 3-4 when not specifying a higher limit while trying to avoid user stories with high similarity if increased to 15, for example. Then, we provide the abstract. \par
\subsection{Evaluation strategy.} We conducted a comprehensive evaluation by utilizing the Quality User Stories (QUS) framework proposed by Lucassen et al.\cite{lucassen2016improving}. The QUS framework provides $13$ quality criteria for assessing the quality of user stories, grouped into three categories: syntactic, semantic, and pragmatic. The quality assessment taxonomy is shown in Table~\ref{tab:my-table}. Some of those criteria are related to the individual user stories, and some are related to the relation of the user story to other user stories in the same set. We note here that we only evaluate semantic and pragmatic quality criteria if the user story is conceptually sound. Moreover, set-related quality criteria (e.g, conflict, uniqueness, .. etc) is evaluated with respect to the user stories from the same abstract and generated by the same LLM.The QUS framework is also applied in ~\cite{10.1007/978-3-031-62110-9_1} to evaluate user stories generated from stakeholder interviews. As no previous work was conducted on this type of work, we use a preliminary experiment conducted using ChatGPT-4 to generate user stories for ten scholarly abstracts as our baseline.\par
In addition to quality assessment, we annotate the user stories for implications of ethics principles following Jobins et al. taxonomy~\cite{Jobin2019}, and non-functional requirements (NFRs) for AI applications identified by Habibullah1 et al.~\cite{Habibullah2023}. For NFRs, we only consider non-ethics-related requirements in the analysis to avoid duplication.\par
The first author and a software engineering graduate student, evaluated the user stories regarding quality attributes, implied ethics principles, and NFRs. Then, to validate the evaluation, 100 user stories were randomly selected and re-evaluated by three software engineers with five years of research and industry experience (No overlap in the re-evaluated attributes). Cohen's Kappa Coefficient was used to calculate the inter-annotator agreement as follows: $\kappa = \frac{P_o - P_e}{1 - P_e}$, where $P_o$ is the observed agreement, and $P_e$ is the expected agreement.

\section{Results}
\label{r}
In this section, we present the empirical evaluation of the generated dataset that will guide us in answering our research question and provide insight into the generated dataset.

\begin{figure}
    \centering
    \includegraphics[width=0.8\linewidth]{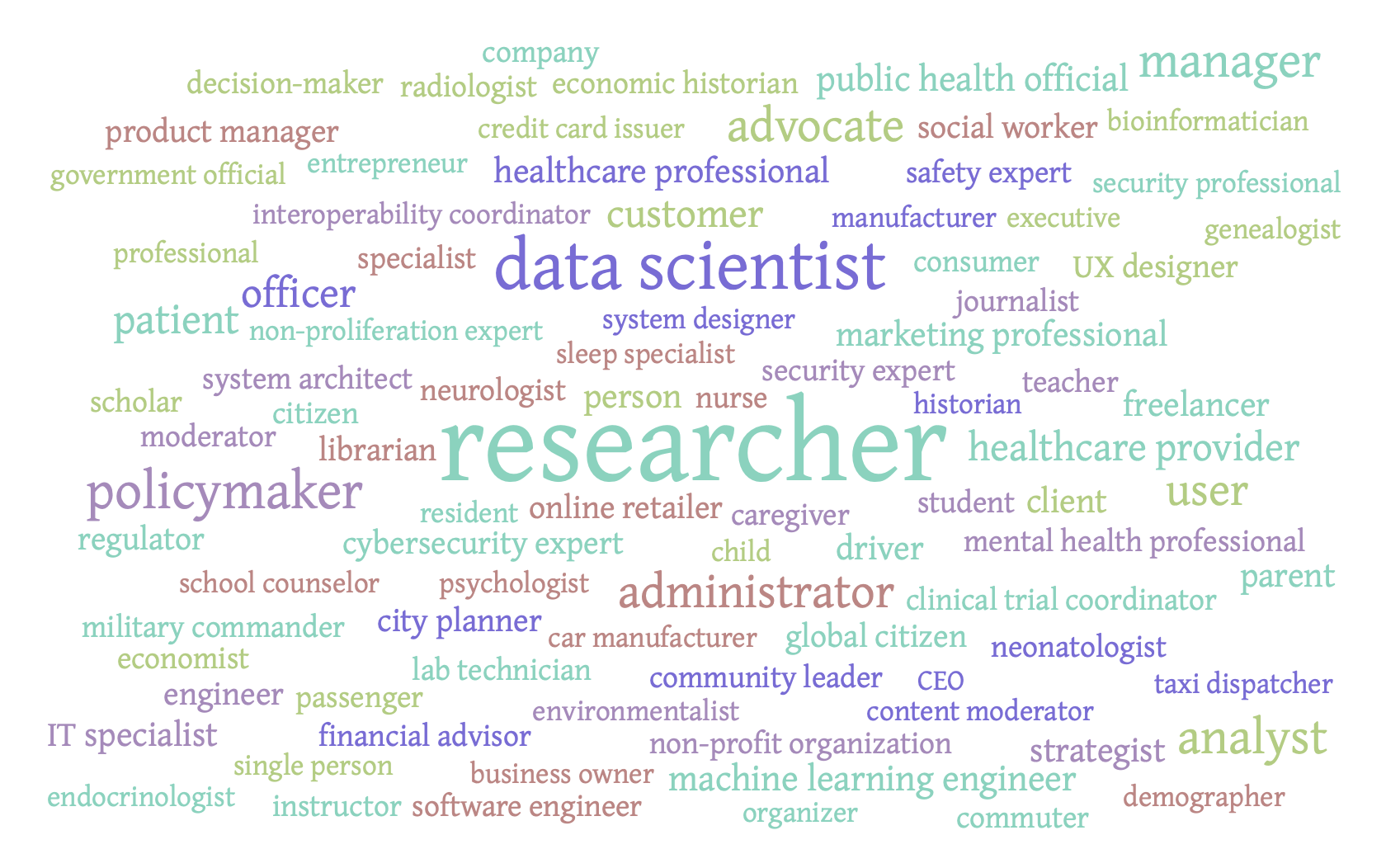}
    \caption{Cloud of words for the generated roles.}
    \label{fig:enter-label}
\end{figure}
\begin{table}[ht]
\centering
\begin{tabular}{lrrrrrr}
\hline
LLM     & min & max & avg & stdev \\
\hline
Gemini  & 14              & 46              & 27.20          & 4.81   \\
Llama   & 19              & 50              & 31.51          & 4.72               \\
O1-mini & 20              & 40              & 27.98          & 3.47                \\
\hline
\end{tabular}
\caption{User story word count statistics per LLM.}
\label{tab:user_story_stats}
\end{table}

\subsection{User stories generation.} A total of $3000$ user stories were generated, out of which $1260$ user stories from $42$ abstracts where selected randomly for analysis. The domain breakdown of the analyzed domains is in the appendix~\ref{appen1}.  A sample of the generated user stories is in Table~\ref{tab:a2-userstories}. The shortest user story generated was generated by Gemini-1.5 flash, while the longest was generated by Llama 3.1 70b. The average user story word count is ~27 for Gemini and O1-mini and 31 for Llama generated user stories. The average user story word count for Llama generated user story is considered higher than user stories collected in~\cite{Dalpiaz2018} from different sources as the average lengths in ~\cite{Dalpiaz2018} ranged between 15-28 by source. More on the dataset statistics is in Table~\ref{tab:user_story_stats}. \par 
The diverse stakeholder representation is embodied in the diverse role clause stated by the user stories. After examining all generated user stories, we explore that all roles reflect various stakeholders relevant to the proposed model in the abstract based on clause 5.3.2 of
the ISO 26000:2010 standard on social responsibility~\cite{iso26000}. However, they focused on individual roles as categorized by~\cite{10.1145/3514094.3534187}, which include users, developers,
engineers, researchers, AI experts, non-users of AI systems, and non-AI experts. No user stories were related to professional bodies such as ACM and IEEE, while seven user stories were related to regulators. Many user stories indicated very specific roles (e.g, child with a history of aggression, healthcare staff member, Robot Maintenance Technician). So, to calculate the identified roles per LLM, we aggregated relevant roles conventionally (e.g., software developer and web developer are mapped to the developer). The roles identified after aggregations per LLM are 113, 91, and 107 for Gemini, Llama, and o1-mini, respectively.

\par
\subsection{Dataset annotation and evaluation}
\textbf{Quality Assessment.} The quality assessment results are detailed in Table~\ref{tab:my-table2}. In terms of syntactic issues, all user stories were \emph{Well-formed}. \emph{O1-mini had the least syntactic issues}, followed by Llama then Gemini. Gemini's main issue was its frequent use of conjunctions, leading to non-atomic user stories.\par In terms of semantic issues, most of the user stories were \emph{Conceptually sound}. Although not counted towards issues under the conceptually sound criteria, 35\% of Llama user stories and 27\% of O1-mini user stories were non-user-centric roles (e.g., researcher, developer, analyst). Gemini had the most problem-oriented user stories, while more than a third of Llama and O1-mini were solution-oriented, which we attribute to the non-user-centric roles. About a third of generated user stories across LLMs are ambiguous. As for conflicts between user stories, Llama and o1-mini had 10\% less conflicts between user stories than Gemini. \emph{Overall, Gemini had the least semantic quality issues followed by O1-mini, then Llama.}\par For pragmatic issues,  all generated user stories were Full sentences and Uniform with extremely minor discrepancies. On the other hand, none of the user story sets were complete. Estimatability was an issue across LLMs, as one-third of user stories were not estimable, which we mostly attribute to ambiguity issues. Independence was the quality criterion violated across LLMs, with more than half of the user stories being dependent, with Gemini being the LLM violating this criterion the most, as many of its user stories involve a mean-end relation across various stakeholders. Lack of uniqueness is another key issue that causes dependency, with Llama and o1-mini having more than 10\% less unique stories than Gemini. \emph{Overall, Gemini had the least pragmatic issues followed by O1-mini, then Llama.}\par
\begin{table*}[!t]
\centering
\caption{Quality Assessment.}
\label{tab:my-table2}
\resizebox{\textwidth}{!}{%
\begin{tabular}{p{0.2\linewidth}p{0.05\linewidth}p{0.06\linewidth}p{0.10\linewidth}p{0.13\linewidth}p{0.10\linewidth}p{0.06\linewidth}p{0.08\linewidth}p{0.08\linewidth}p{0.05\linewidth}p{0.10\linewidth}}

\toprule
Quality Criteria & Atomic &	Minimal &	Conceptually sound	& Problem-oriented	& Unambiguous &	Conflict-free	& Estimatable	& Independent &	Unique & Mean-End Relation\\ \midrule
Gemini 1.5-flash   & 75\%  & 93\% & 85\% & 83\% & 62\% & 85\%  & 68\% & 39\% & 78\% & 47\% \\
Llama 3.1 70b   & 83\%  & 89\% & 88\% & 88\% & 64\% & 96\%  & 64\% & 34\% & 67\% & 45\% \\
o1-mini   & 90\%  & 98\% & 84\% & 84\% & 61\% & 96\%  & 64\% & 46\% & 64\% & 40\% \\
GPT-4 (preliminary exp.) & 62\% & 83\% & 70\%  & 79\% & 30\% & - & -& - & - & - \\

\bottomrule
\end{tabular}%

}
\end{table*}

\begin{figure*}[!h]
    \centering
    \includegraphics[width=0.8\linewidth]{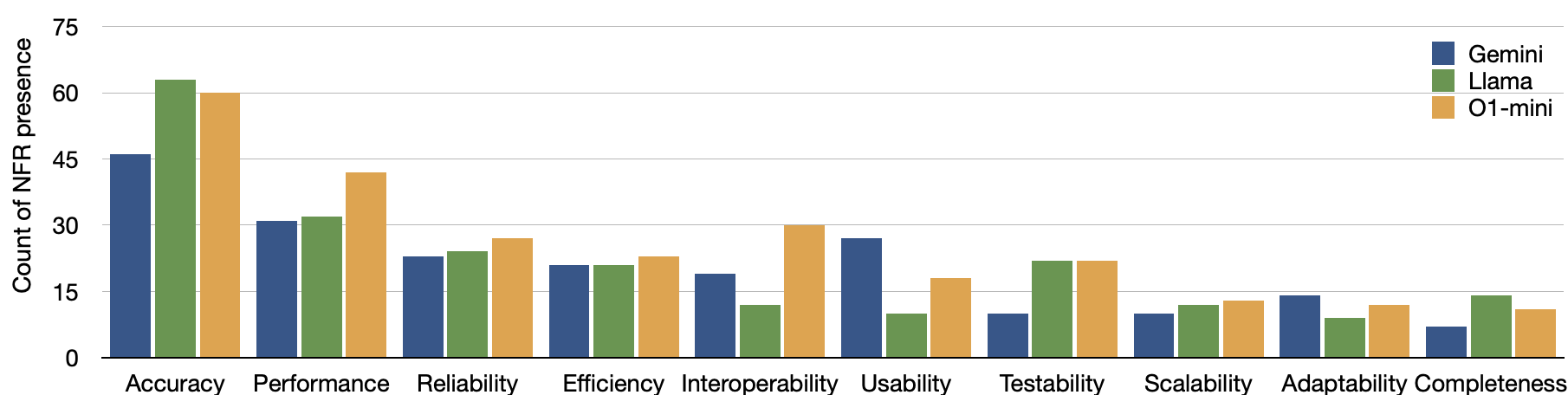}
    \caption{Occurrence of NFR per LLM (top 10 only, ethics related NFRs excluded).}
    \label{fig:nfr}
\end{figure*}

\textbf{Non-Functional Requirements.} The generated requirements put great emphasis on NFRs, as 55\% of the conceptually sound generated user stories emphasize at least one NFR. All investigated LLMs agreed on accuracy and performance as the most important NFRs,as shown in Figure~\ref{fig:nfr} . Reliability and efficiency were also among the top 5 mentioned NFRs. Major differences are that Llama favored testability, O1-mini favored interoperability, while Gemini emphasizes usability.

\begin{figure*}[!h]
    \centering
    \includegraphics[width=0.8\linewidth]{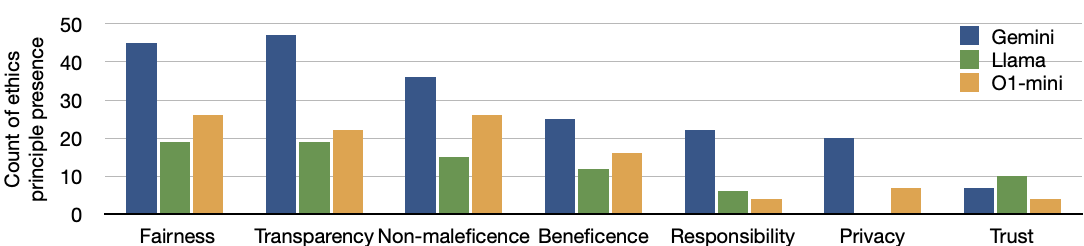}
    \caption{Occurrence of ethics principles per LLM (top 7 only).}
    \label{fig:ethics}
\end{figure*}

\textbf{Ethics requirements.}
Gemini was the most sensitive to ethics requirements when generating user stories, as ethics principles were mentioned or implied in 33\%  of user stories covering 9 out of 11 ethics principles in~\cite{Jobin2019}. As for Llama and o1-mini, they enlisted ethics principles in 15\% and 20\% in their user stories, respectively. The three most important ethical principles mentioned or implied are justice and fairness, transparency, and non-maleficence. Details on the number of each ethics requirement are mentioned in Figure~\ref{fig:ethics}.  

\subsection{Annotation validation.} Ten user story sets, which consist of 100 user stories, and ~8\% of the generated user stories were re-evaluated by three software engineers with $5$ years of industry experience. The $P_o$ was $0.965$, $0.94$, and $0.88$ for the quality attributes, ethical principles, and NFRs, respectively. $P_e = 0.50$, assuming equal probability for agreement and disagreement. Therefore, $\kappa_{quality} = 0.93$, $\kappa_{ethics} = 0.88$, indicating an "almost perfect" inter-annotator agreement for QUS attributes and ethical principles. Whereas $\kappa_{NFR} = 0.76$ indicates a substantial inter-annotator agreement for NFRs.


\begin{figure*}[h]
    \centering
    \includegraphics[width=0.65\linewidth]{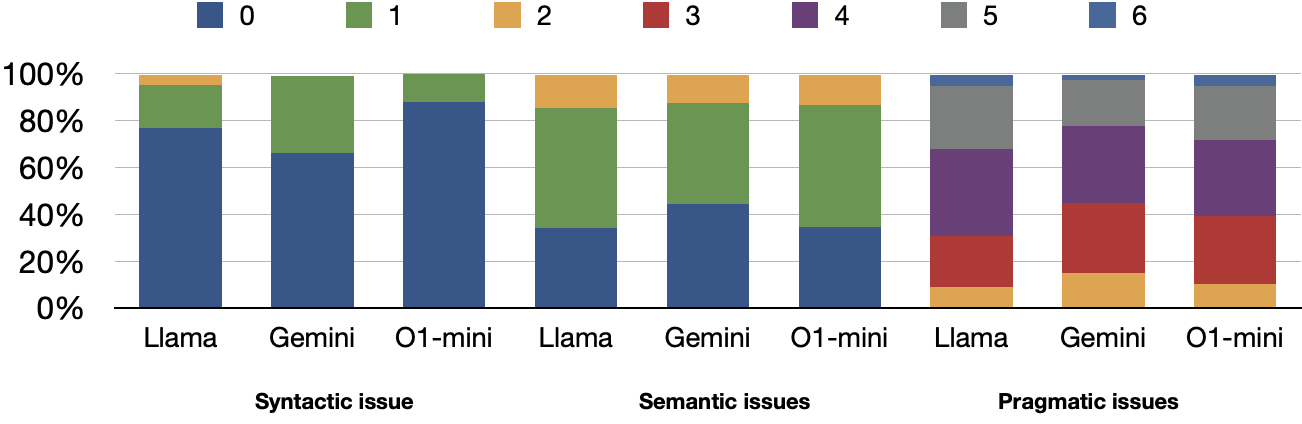}
    \caption{Percentage of user stories with quality issues.}
    \label{fig:qs}
\end{figure*}
\section{Discussion}
\label{d}
In this section, we answer our research questions and elaborate on some remarks.
\subsection{Can LLMs be used to generate user stories from the perspective of multiple stakeholders based on the abstract description of the AI component?}
\label{rq1}
Yes, as demonstrated in the results section, \emph{LLMs are capable of generating user stories for a diverse set of stakeholders.} In a zero-shot learning context, the models successfully create user stories that extend beyond the abstract's focus on solutions and outcomes to address NFRs, including ethics requirements. For example, for an AI system for Aggression Detection in Children with ADHD the Gemini LLM generates the user story A2US9Ge: \enquote{As a data privacy advocate, I want assurances that any wearable device used for aggression detection collects data securely and protects children's privacy.} While the abstract does not mention privacy as a feature or concern, a user story was generated with this regard, highlighting the usefulness of Gemini in the early phases of requirements engineering.

\subsection{Which LLM generates the highest quality user stories according to the QUS framework \cite{lucassen2016improving}?}
\label{RQ2}
To answer this question, we assess each aspect shown in Figure~\ref{fig:qs}.\\

\textbf{Syntactic aspect.}
Starting with the \emph{syntactic aspect}, as mentioned in the results, O1-mini produced user stories that were more syntactically sound. However, beyond the requirement for a user story to be well-formed, syntactic quality issues, such as the presence of brackets or examples, are easier to resolve. The main cause of syntactic issues in stories generated by both LLMs, conjunctions, can be addressed by splitting user stories into multiple simpler ones.\\

\textbf{Semantic aspect.}
As for the \emph{semantic aspect}, Gemini generated 10\% more user stories with no semantic quality issues. Ambiguity was a common problem across all LLMs, often caused by vagueness in measuring NFRs or the generalities in user stories. Llama and O1-mini user stories frequently exhibited solution-oriented perspectives, driven by non-user-centric roles, contrary to the user-centric focus required in user stories. User stories should focus on the end-user and stakeholders, not what the developer wants to build unless developers are stakeholders themselves (e.g., developer frameworks). For example, A2US9Ll: \enquote{As a clinician, I want to use the sensor-derived feature of vector magnitude (faster triaxial acceleration) to identify early warning signs of physical-aggressive incidents in children so that I can develop targeted interventions and prevention strategies.}\par
A notable semantic issue in Gemini was the presence of conflict due to its emphasis on ethics-focused requirements. For example, there was a conflict between a privacy advocate in A2US9Ge presented earlier and researchers in A2US4Ge: \enquote{As a researcher studying childhood aggression, I want access to a non-invasive method for collecting real-time data on physical aggression in children so that I can better understand its triggers and consequences,} who sought real-time data on physical aggression in children. Similar conflicts were observed across all LLMs when different stakeholders prioritized conflicting NFRs. For instance, a conflict arose in a healthcare robot delivery system between A5US5Ge: \enquote{As a patient, I want to receive my meals promptly and accurately delivered by the robot to enhance my hospital stay,} where patients prioritized delivery speed, and A5US8Ge: \enquote{As a facility manager, I want to ensure the robot operates safely and securely within the facility,} where the facility manager emphasized safety and security. These examples highlight the need for acceptance criteria that balance NFRs effectively.\\

\textbf{Pragmatic aspect.}
In the \emph{pragmatic aspect}, Gemini generated a higher percentage of user stories with three or fewer pragmatic issues, outperforming O1-mini by 5\% and Llama by 14\%. The most prevalent pragmatic issue was high dependency among user stories, mainly due to their reliance on foundational tasks for ML systems (e.g., dataset construction, model training, and evaluation). Additionally, semantic duplication and repetitive roles contributed to dependency issues, with a third of Llama and O1-mini user stories being duplicates or involving different roles with the same means or ends. Non-user-centric roles also led to more \emph{End = Means} relationships, where non-user-centric stories acted as means to user-centric ends. Hence, we find this issue mainly in Llama-generated stories. An example would be in the case of A10US9Ll: \enquote{As a researcher, I want to improve the precision of the predictive model so that I can increase the confidence of patients and healthcare professionals in the system} and A10US10Ll: \enquote{As a person with diabetes, I want to use a predictive model that can help me avoid false alarms and unnecessary interventions so that I can manage my blood glucose levels more effectively and improve my quality of life,} with A10US9 being a means and A10US10 an end.\par

Estimating the generated user stories was further complicated by high uncertainty in the research processes required for ML development and vagueness in NFRs, particularly in sensitive contexts requiring high precision or accuracy. Generality also posed challenges, as seen in A7US9Ge: \enquote{As a developer of the mobile app for the tool, I want to develop a user-friendly interface that is culturally sensitive and inclusive for diverse target groups.} or in A1US2Ge: \enquote{As a city planner, I want a traffic management system that uses AI and connected vehicles to optimize traffic flow and reduce congestion.}  Addressing these issues often required introducing acceptance criteria or breaking stories into sub-stories. Lastly, the lack of completeness in user story sets was attributed to the limited information in scholarly paper abstracts and the restriction of prompts to only ten user stories. \emph{Overall, Gemini exhibited fewer quality issues compared to O1-mini and Llama-generated user stories.}\par

Comparing to GPT‑4‑generated user stories, shown in Table \ref{tab:my-table2}, we can see the significant improvements when using the more advanced LLMs. The generated user stories from Gemini 1.5‑Flash, Llama 3.1 70 B, and o1‑mini are consistently more atomic, minimal, conceptually sound, and unambiguous than the earlier GPT‑4 baseline.However, when compared to a similar study in which practitioners crafted user stories based on interviews \cite{10.1007/978-3-031-62110-9_1}, it becomes evident that human-generated stories still set the benchmark for quality. These interview-derived stories outperformed all LLM-generated counterparts in atomicity, minimality, and conceptual soundness, with scores exceeding 88\%. Nonetheless, estimatability and ambiguity remain persistent challenges for both LLM- and human-generated stories, as even the human-crafted set achieved only 77\% in these aspects.



\subsection{To what extent do LLM-generated user stories capture ethical considerations and principles and NFRs relevant to AI systems?
}
\emph{LLM generated user stories do capture a diverse array of NFRs and ethical principles}, even when such requirements or principles were not referenced in the abstract, as discussed earlier when answering RQ1~\ref{rq1}. Different stakeholders, encoded in the role, demand different NFRs. While user-centric and non-user-centric roles highly emphasize accuracy and performance, user-centric roles highly value reliability, efficiency, and usability. On the other hand, non-user-centric roles emphasized interoperability, testability, and scalability, demonstrating that stakeholders are more sensitive to factors that impact their immediate interaction with the system. It also highlights the importance of prioritization and conflict resolution as conflict was a main semantic issue, as discussed earlier when answering RQ1~\ref{RQ2}. It's worth noting that not all NFRs regarded as highly important by surveyed participants from academic and industry backgrounds in~\cite{Habibullah2023} were mentioned by the generated user stories as reproducibility and traceability.

As for ethical principles, the top three ethical principles implied or mentioned in the user stories are also the top three identified in surveyed documents in Jobins et al.~\cite{Jobin2019} survey. When considering the alignment between the Jobins et al.~\cite{Jobin2019} survey and the LLM's ranking in terms of the frequency in implying an ethical principle the Spearman rho is 0.936, 0.841, 0.753 (p < 0.005) for Gemini, O1-mini, and Llama, respectively. While there is a high alignment, as noted in~\cite{bano2023aialloperationalisingdiversity}, many user stories on ethics lack the depth needed and require further refinements to remove ambiguities.\par

\subsection{Improving the quality of generated user stories.} 
You can improve the quality of the generated user stories by prompting the LLM to do so. A prompt such as "Remove conjunctions in <user story> by splitting the user story" removes the conjunction and creates a user story for each feature. Even more subjective qualities, such as ambiguity, can be addressed similarly. Asking Gemini to remove ambiguity from user story A7US9Ge, mentioned earlier, resulted in a user story that was less ambiguous but not atomic: "As a developer of the mobile app for the tool, I want to design an interface that is clear, intuitive, and accessible, with language and imagery that respects diverse cultural backgrounds. The app should support multiple languages and offer customization options to accommodate users with varying levels of technological literacy, mental health conditions, and cultural preferences. This will ensure that the app is inclusive and easy to use for all target groups." To resolve the atomicity issues, we further asked it to break it down into atomic stories, which resulted in a total of $11$ less ambiguous and atomic stories such as: "As a user who speaks [specific language], I want the app to be available in my native language so that I can understand and use it comfortably," under the Multilingual Support theme, and "As a user with visual impairments, I want the app to support screen readers and have adjustable font sizes and color contrast options so that I can use it effectively," as well as "As a user with motor impairments, I want the app to have large touch targets and support alternative input methods (e.g., voice commands) so that I can interact with it easily,"  under the Accessibility theme. \par Upon further experimentation, we found that addressing issues after generation is more effective than focusing on them during the initial prompt, as early focus on such details can result in less holistic initial sets in order to accommodate attributes such as ambiguous or independent.\par

\subsection{Dataset usage} This data set can be used in various ways including, but not limited to:\\
\textbf{1- Human-centric research related to Requirements in AI systems} UStAI can be used to investigate ethical issues and concerns in various domains and as a seed dataset to elicit ethics-related requirements from the general population. Given that the dataset includes annotations for ethical principles for user stories that explicitly indicate ethical requirements, it can be used for AI-ethics-related research such as prioritization of ethics requirements, conflict resolution between ethical priorities for different stakeholders, and training models to detect ethical issues in software requirements documents. Moreover, as 55\% of the user stories mention or imply at least one NFR, UStAI can be used to explore methods for automatically extracting and prioritizing NFRs in AI-based systems.\\
\textbf{2- Benchmarking and Evaluation of LLMs for Requirements Engineering} Researchers can use UStAI as a benchmark to compare LLMs in their ability to generate high-quality user stories based on the QUS framework with the quality of user stories generated by the three LLMs we investigated. Moreover, this dataset offers a basis for experimenting with various prompt engineering techniques for user story generation, leading to more effective ways of eliciting requirements from LLMs. In addition, UStAI can be used to investigate LLM's ability to complete a set of requirements given a set of high-level requirements.\\
\textbf{3- Training and Evaluation of NLP techniques:} As UStAI is annotated with quality metrics, it can be used as training data for developing quality assessment tools using efficient and lightweight techniques. Also, it can be used as an evaluation dataset for existing methods and techniques for conflict detection, ambiguity detection, and other requirements smells. UStAI could be further annotated to be usable for other requirements engineering tasks.

\subsection{Implications and future directions} The results and discussion demonstrate that utilizing research and technical paper summaries provides a viable alternative for exploring requirements during the early phases of Requirements Engineering (RE). The key issues observed in Gemini mainly result from other favorable attributes, such as its emphasis on ethical principles and the diversity of end-user stakeholders. However, it is important to recognize the potential value of different LLMs in various contexts. For instance, most quality issues in O1-mini and Llama-generated user stories stem from their faithfulness to the abstract (the source of the generated user stories). This makes O1-mini and Llama preferable in scenarios where high precision is required. Additionally, Llama, being an open-source LLM, presents an optimal choice for on-premise solutions.\par
However, current LLMs are not yet ready to fully support or replace human involvement in RE tasks, highlighting several promising areas for future research. For example, exploring alternative sources for generating user stories beyond scholarly abstracts, such as issue trackers or industrial reports. Additionally, further investigation is required to understand the performance variations among different LLMs in generating user stories, particularly regarding relevance in terms of scope, roles, and ethical considerations, as well as the alignment between ethics-related user stories and stakeholder values. Moreover, agentic AI could be employed to enhance completeness and reduce ambiguities, especially for non-functional and ethics-related requirements. Another valuable direction is the empirical evaluation of human-in-the-loop approaches for eliciting domain-specific requirements requiring detailed expertise. While this paper rigorously evaluates generated user stories using the QUS framework, future studies would benefit from validation involving real-world AI development teams or industry practitioners to assess practical applicability.

\subsection{Threats to validity}

\textit{Internal Validity:} Abstract selection bias is a potential concern. To mitigate this, 42 abstracts were randomly selected from 28 domains, ensuring diversity in quality and authorship. Another potential threat was related to prompt design; however, this was addressed through an exploration phase to refine and optimize the prompt.
\textit{External Validity:} Since LLMs evolve in a fast pace results may change over time or not apply to other LLMs from the same family. \textit{Conclusion Validity:} Quality assessment reliability is a potential threat. This was mitigated by involving two annotators in the evaluation process and having 100 of the user stories (~8\%) re-evaluated by three software engineer with industry experience, ensuring consistency and credibility in the assessments. 
\par

\section{Conclusion}
\label{c}
This work contributes to the growing body of knowledge on applied LLMs and the field of RE4AI. We proposed and evaluated generating user stories from abstract descriptions to build a synthetic AI user story dataset annotated with quality attributes, NFRs, and implied ethical principles. Our research demonstrated the potential of LLMs in producing high-quality user stories from abstract descriptions of AI components. It identified strengths and limitations in the syntactic, semantic, and pragmatic aspects and analyzed user stories for the presence of NFRs and ethical principles. The empirical evaluation concluded that Gemini 1.5 Flash outperformed other investigated LLMs by generating stories with fewer quality issues, greater emphasis on ethics, and a broader perspective, making it the preferred tool for early-stage requirements elicitation. We release the UStAI dataset of $1260$ manually annotated user stories and $1740$ generated non-annotated user stories to the research community. Future work will explore automating quality evaluation and ethical alignment to advance tools designed to assist in eliciting ethical software requirements for AI systems.

\section*{Acknowledgment}
The authors would like to acknowledge the support received from Saudi Data and AI Authority (SDAIA) and King Fahd University of Petroleum and Minerals (KFUPM) under SDAIA-KFUPM Joint Research Center for Artificial Intelligence Grant no. JRC-AI-RG-12. The authors would also like to thank Eman Alofi, Nadeen Alamoudi, Mais Al-heraki, and Linah Abouhajar for their contribution to the annotation and validation.

\bibliographystyle{ACM-Reference-Format}
\bibliography{custom}


\begin{thebibliography}{33}


\ifx \showCODEN    \undefined \def \showCODEN     #1{\unskip}     \fi
\ifx \showISBNx    \undefined \def \showISBNx     #1{\unskip}     \fi
\ifx \showISBNxiii \undefined \def \showISBNxiii  #1{\unskip}     \fi
\ifx \showISSN     \undefined \def \showISSN      #1{\unskip}     \fi
\ifx \showLCCN     \undefined \def \showLCCN      #1{\unskip}     \fi
\ifx \shownote     \undefined \def \shownote      #1{#1}          \fi
\ifx \showarticletitle \undefined \def \showarticletitle #1{#1}   \fi
\ifx \showURL      \undefined \def \showURL       {\relax}        \fi
\providecommand\bibfield[2]{#2}
\providecommand\bibinfo[2]{#2}
\providecommand\natexlab[1]{#1}
\providecommand\showeprint[2][]{arXiv:#2}

\bibitem[Abed et~al\mbox{.}(2024)]%
        {10.1007/978-3-031-62110-9_1}
\bibfield{author}{\bibinfo{person}{Omed Abed}, \bibinfo{person}{Karsten Nebe}, {and} \bibinfo{person}{Ahmed~Belal Abdellatif}.} \bibinfo{year}{2024}\natexlab{}.
\newblock \showarticletitle{AI-Generated User Stories Supporting Human-Centred Development: An Investigation on Quality}. In \bibinfo{booktitle}{\emph{HCI International 2024 Posters}}, \bibfield{editor}{\bibinfo{person}{Constantine Stephanidis}, \bibinfo{person}{Margherita Antona}, \bibinfo{person}{Stavroula Ntoa}, {and} \bibinfo{person}{Gavriel Salvendy}} (Eds.). \bibinfo{publisher}{Springer Nature Switzerland}, \bibinfo{address}{Cham}, \bibinfo{pages}{3--13}.
\newblock
\showISBNx{978-3-031-62110-9}


\bibitem[Bano et~al\mbox{.}(2023)]%
        {bano2023aialloperationalisingdiversity}
\bibfield{author}{\bibinfo{person}{Muneera Bano}, \bibinfo{person}{Didar Zowghi}, \bibinfo{person}{Vincenzo Gervasi}, {and} \bibinfo{person}{Rifat Shams}.} \bibinfo{year}{2023}\natexlab{}.
\newblock \bibinfo{title}{AI for All: Operationalising Diversity and Inclusion Requirements for AI Systems}.
\newblock
\showeprint[arxiv]{2311.14695}~[cs.CY]
\urldef\tempurl%
\url{https://arxiv.org/abs/2311.14695}
\showURL{%
\tempurl}


\bibitem[Boetticher(2007)]%
        {boetticher2007promise}
\bibfield{author}{\bibinfo{person}{G Boetticher}.} \bibinfo{year}{2007}\natexlab{}.
\newblock \showarticletitle{The PROMISE repository of empirical software engineering data}.
\newblock \bibinfo{journal}{\emph{http://promisedata. org/repository}} (\bibinfo{year}{2007}).
\newblock


\bibitem[Cohn(2004)]%
        {Cohn2004-zk}
\bibfield{author}{\bibinfo{person}{Mike Cohn}.} \bibinfo{year}{2004}\natexlab{}.
\newblock \bibinfo{booktitle}{\emph{User stories applied}}.
\newblock \bibinfo{publisher}{Addison-Wesley Educational}, \bibinfo{address}{Boston, MA}.
\newblock


\bibitem[Dalpiaz(2018)]%
        {Dalpiaz2018}
\bibfield{author}{\bibinfo{person}{Fabiano Dalpiaz}.} \bibinfo{year}{2018}\natexlab{}.
\newblock \bibinfo{title}{Requirements data sets (user stories)}.
\newblock
\href{https://doi.org/10.17632/7zbk8zsd8y.1}{doi:\nolinkurl{10.17632/7zbk8zsd8y.1}}
\newblock
\shownote{Data set}.


\bibitem[Deshpande and Sharp(2022)]%
        {10.1145/3514094.3534187}
\bibfield{author}{\bibinfo{person}{Advait Deshpande} {and} \bibinfo{person}{Helen Sharp}.} \bibinfo{year}{2022}\natexlab{}.
\newblock \showarticletitle{Responsible AI Systems: Who are the Stakeholders?}. In \bibinfo{booktitle}{\emph{Proceedings of the 2022 AAAI/ACM Conference on AI, Ethics, and Society}} (Oxford, United Kingdom) \emph{(\bibinfo{series}{AIES '22})}. \bibinfo{publisher}{Association for Computing Machinery}, \bibinfo{address}{New York, NY, USA}, \bibinfo{pages}{227–236}.
\newblock
\showISBNx{9781450392471}
\href{https://doi.org/10.1145/3514094.3534187}{doi:\nolinkurl{10.1145/3514094.3534187}}


\bibitem[Dwitama and Rusli(2020)]%
        {Dwitama2020}
\bibfield{author}{\bibinfo{person}{Ferliana Dwitama} {and} \bibinfo{person}{Andre Rusli}.} \bibinfo{year}{2020}\natexlab{}.
\newblock \showarticletitle{User stories collection via interactive chatbot to support requirements gathering}.
\newblock \bibinfo{journal}{\emph{TELKOMNIKA (Telecommunication Computing Electronics and Control)}} \bibinfo{volume}{18}, \bibinfo{number}{2} (\bibinfo{date}{April} \bibinfo{year}{2020}), \bibinfo{pages}{890}.
\newblock
\showISSN{1693-6930}
\href{https://doi.org/10.12928/telkomnika.v18i2.14866}{doi:\nolinkurl{10.12928/telkomnika.v18i2.14866}}


\bibitem[Fabris et~al\mbox{.}(2022)]%
        {Fabris2022}
\bibfield{author}{\bibinfo{person}{Alessandro Fabris}, \bibinfo{person}{Stefano Messina}, \bibinfo{person}{Gianmaria Silvello}, {and} \bibinfo{person}{Gian~Antonio Susto}.} \bibinfo{year}{2022}\natexlab{}.
\newblock \showarticletitle{Algorithmic fairness datasets: the story so far}.
\newblock \bibinfo{journal}{\emph{Data Mining and Knowledge Discovery}} \bibinfo{volume}{36}, \bibinfo{number}{6} (\bibinfo{date}{Sept.} \bibinfo{year}{2022}), \bibinfo{pages}{2074–2152}.
\newblock
\showISSN{1573-756X}
\href{https://doi.org/10.1007/s10618-022-00854-z}{doi:\nolinkurl{10.1007/s10618-022-00854-z}}


\bibitem[Ferrara et~al\mbox{.}(2024)]%
        {10.1145/3597503.3639185}
\bibfield{author}{\bibinfo{person}{Carmine Ferrara}, \bibinfo{person}{Francesco Casillo}, \bibinfo{person}{Carmine Gravino}, \bibinfo{person}{Andrea De~Lucia}, {and} \bibinfo{person}{Fabio Palomba}.} \bibinfo{year}{2024}\natexlab{}.
\newblock \showarticletitle{ReFAIR: Toward a Context-Aware Recommender for Fairness Requirements Engineering}. In \bibinfo{booktitle}{\emph{Proceedings of the IEEE/ACM 46th International Conference on Software Engineering}} (Lisbon, Portugal) \emph{(\bibinfo{series}{ICSE '24})}. \bibinfo{publisher}{Association for Computing Machinery}, \bibinfo{address}{New York, NY, USA}, Article \bibinfo{articleno}{213}, \bibinfo{numpages}{12}~pages.
\newblock
\showISBNx{9798400702174}
\href{https://doi.org/10.1145/3597503.3639185}{doi:\nolinkurl{10.1145/3597503.3639185}}


\bibitem[Ferrari et~al\mbox{.}(2017)]%
        {Ferrari2017PURE}
\bibfield{author}{\bibinfo{person}{Alessio Ferrari}, \bibinfo{person}{Giorgio~Oronzo Spagnolo}, {and} \bibinfo{person}{Stefania Gnesi}.} \bibinfo{year}{2017}\natexlab{}.
\newblock \showarticletitle{PURE: A {Dataset} of {Public} {Requirements} {Documents}}. In \bibinfo{booktitle}{\emph{2017 {IEEE} 25th {International} {Requirements} {Engineering} {Conference} ({RE})}}. IEEE, \bibinfo{pages}{502--505}.
\newblock
\href{https://doi.org/10.1109/re.2017.29}{doi:\nolinkurl{10.1109/re.2017.29}}


\bibitem[Habibullah et~al\mbox{.}(2023)]%
        {Habibullah2023}
\bibfield{author}{\bibinfo{person}{Khan~Mohammad Habibullah}, \bibinfo{person}{Gregory Gay}, {and} \bibinfo{person}{Jennifer Horkoff}.} \bibinfo{year}{2023}\natexlab{}.
\newblock \showarticletitle{Non-functional requirements for machine learning: understanding current use and challenges among practitioners}.
\newblock \bibinfo{journal}{\emph{Requirements Engineering}} \bibinfo{volume}{28}, \bibinfo{number}{2} (\bibinfo{date}{Jan.} \bibinfo{year}{2023}), \bibinfo{pages}{283–316}.
\newblock
\showISSN{1432-010X}
\href{https://doi.org/10.1007/s00766-022-00395-3}{doi:\nolinkurl{10.1007/s00766-022-00395-3}}


\bibitem[Halme et~al\mbox{.}(2024)]%
        {Halme2024}
\bibfield{author}{\bibinfo{person}{Erika Halme}, \bibinfo{person}{Marianna Jantunen}, \bibinfo{person}{Ville Vakkuri}, \bibinfo{person}{Kai-Kristian Kemell}, {and} \bibinfo{person}{Pekka Abrahamsson}.} \bibinfo{year}{2024}\natexlab{}.
\newblock \showarticletitle{Making ethics practical: User stories as a way of implementing ethical consideration in Software Engineering}.
\newblock \bibinfo{journal}{\emph{Information and Software Technology}}  \bibinfo{volume}{167} (\bibinfo{date}{March} \bibinfo{year}{2024}), \bibinfo{pages}{107379}.
\newblock
\showISSN{0950-5849}
\href{https://doi.org/10.1016/j.infsof.2023.107379}{doi:\nolinkurl{10.1016/j.infsof.2023.107379}}


\bibitem[Huffman~Hayes et~al\mbox{.}(2018)]%
        {Huffman2018REquirements}
\bibfield{author}{\bibinfo{person}{Jane Huffman~Hayes}, \bibinfo{person}{Alex Dekhtyar}, {and} \bibinfo{person}{Jared Payne}.} \bibinfo{year}{2018}\natexlab{}.
\newblock \showarticletitle{The {REquirements} {TRacing} {On} {Target} ({RETRO}).{NET} {Dataset}}. In \bibinfo{booktitle}{\emph{2018 {IEEE} 26th {International} {Requirements} {Engineering} {Conference} ({RE})}}. IEEE, \bibinfo{pages}{424--427}.
\newblock
\href{https://doi.org/10.1109/re.2018.00054}{doi:\nolinkurl{10.1109/re.2018.00054}}


\bibitem[Inayat et~al\mbox{.}(2015)]%
        {10.1016/j.chb.2014.10.046}
\bibfield{author}{\bibinfo{person}{Irum Inayat}, \bibinfo{person}{Siti~Salwah Salim}, \bibinfo{person}{Sabrina Marczak}, \bibinfo{person}{Maya Daneva}, {and} \bibinfo{person}{Shahaboddin Shamshirband}.} \bibinfo{year}{2015}\natexlab{}.
\newblock \showarticletitle{A systematic literature review on agile requirements engineering practices and challenges}.
\newblock \bibinfo{journal}{\emph{Comput. Hum. Behav.}} \bibinfo{volume}{51}, \bibinfo{number}{PB} (\bibinfo{date}{oct} \bibinfo{year}{2015}), \bibinfo{pages}{915–929}.
\newblock
\showISSN{0747-5632}
\href{https://doi.org/10.1016/j.chb.2014.10.046}{doi:\nolinkurl{10.1016/j.chb.2014.10.046}}


\bibitem[{ISO/TMBG Technical Management Board - groups}(2010)]%
        {iso26000}
\bibfield{author}{\bibinfo{person}{{ISO/TMBG Technical Management Board - groups}}.} \bibinfo{year}{2010}\natexlab{}.
\newblock \bibinfo{title}{ISO 26000:2010 Guidance on social responsibility}.
\newblock \bibinfo{howpublished}{\url{https://www.iso.org/cms/render/live/en/sites/isoorg/contents/data/standard/04/25/42546.html}}.
\newblock
\newblock
\shownote{Retrieved March 1, 2022}.


\bibitem[Jobin et~al\mbox{.}(2019)]%
        {Jobin2019}
\bibfield{author}{\bibinfo{person}{Anna Jobin}, \bibinfo{person}{Marcello Ienca}, {and} \bibinfo{person}{Effy Vayena}.} \bibinfo{year}{2019}\natexlab{}.
\newblock \showarticletitle{The global landscape of AI ethics guidelines}.
\newblock \bibinfo{journal}{\emph{Nature Machine Intelligence}}  \bibinfo{volume}{1} (\bibinfo{date}{9} \bibinfo{year}{2019}), \bibinfo{pages}{389--399}.
\newblock
Issue 9.
\href{https://doi.org/10.1038/s42256-019-0088-2}{doi:\nolinkurl{10.1038/s42256-019-0088-2}}


\bibitem[Knauss et~al\mbox{.}(2021)]%
        {Knauss2021}
\bibfield{author}{\bibinfo{person}{Eric Knauss}, \bibinfo{person}{Sigrid~H. Houmb}, \bibinfo{person}{Shareeful Islam}, \bibinfo{person}{Jan Jürjens}, {and} \bibinfo{person}{Kurt Schneider}.} \bibinfo{year}{2021}\natexlab{}.
\newblock \bibinfo{title}{SecReq}.
\newblock
\href{https://doi.org/10.5281/zenodo.4530183}{doi:\nolinkurl{10.5281/zenodo.4530183}}
\newblock
\shownote{Data set}.


\bibitem[Kumichev et~al\mbox{.}(2024)]%
        {kumichev2024medsynllmbasedsyntheticmedical}
\bibfield{author}{\bibinfo{person}{Gleb Kumichev}, \bibinfo{person}{Pavel Blinov}, \bibinfo{person}{Yulia Kuzkina}, \bibinfo{person}{Vasily Goncharov}, \bibinfo{person}{Galina Zubkova}, \bibinfo{person}{Nikolai Zenovkin}, \bibinfo{person}{Aleksei Goncharov}, {and} \bibinfo{person}{Andrey Savchenko}.} \bibinfo{year}{2024}\natexlab{}.
\newblock \bibinfo{title}{MedSyn: LLM-based Synthetic Medical Text Generation Framework}.
\newblock
\showeprint[arxiv]{2408.02056}~[cs.CL]
\urldef\tempurl%
\url{https://arxiv.org/abs/2408.02056}
\showURL{%
\tempurl}


\bibitem[Lima et~al\mbox{.}(2019)]%
        {lima2019software}
\bibfield{author}{\bibinfo{person}{M{\'a}rcia Lima}, \bibinfo{person}{Victor Valle}, \bibinfo{person}{Estev{\~a}o Costa}, \bibinfo{person}{Fylype Lira}, {and} \bibinfo{person}{Bruno Gadelha}.} \bibinfo{year}{2019}\natexlab{}.
\newblock \showarticletitle{Software engineering repositories: expanding the promise database}. In \bibinfo{booktitle}{\emph{Proceedings of the XXXIII Brazilian Symposium on Software Engineering}}. \bibinfo{pages}{427--436}.
\newblock


\bibitem[Long et~al\mbox{.}(2024)]%
        {long-etal-2024-llms}
\bibfield{author}{\bibinfo{person}{Lin Long}, \bibinfo{person}{Rui Wang}, \bibinfo{person}{Ruixuan Xiao}, \bibinfo{person}{Junbo Zhao}, \bibinfo{person}{Xiao Ding}, \bibinfo{person}{Gang Chen}, {and} \bibinfo{person}{Haobo Wang}.} \bibinfo{year}{2024}\natexlab{}.
\newblock \showarticletitle{On {LLM}s-Driven Synthetic Data Generation, Curation, and Evaluation: A Survey}. In \bibinfo{booktitle}{\emph{Findings of the Association for Computational Linguistics ACL 2024}}, \bibfield{editor}{\bibinfo{person}{Lun-Wei Ku}, \bibinfo{person}{Andre Martins}, {and} \bibinfo{person}{Vivek Srikumar}} (Eds.). \bibinfo{publisher}{Association for Computational Linguistics}, \bibinfo{address}{Bangkok, Thailand and virtual meeting}, \bibinfo{pages}{11065--11082}.
\newblock
\urldef\tempurl%
\url{https://aclanthology.org/2024.findings-acl.658}
\showURL{%
\tempurl}


\bibitem[Lucassen et~al\mbox{.}(2016)]%
        {lucassen2016improving}
\bibfield{author}{\bibinfo{person}{Garm Lucassen}, \bibinfo{person}{Fabiano Dalpiaz}, \bibinfo{person}{Jan Martijn~EM van~der Werf}, {and} \bibinfo{person}{Sjaak Brinkkemper}.} \bibinfo{year}{2016}\natexlab{}.
\newblock \showarticletitle{Improving agile requirements: the quality user story framework and tool}.
\newblock \bibinfo{journal}{\emph{Requirements engineering}}  \bibinfo{volume}{21} (\bibinfo{year}{2016}), \bibinfo{pages}{383--403}.
\newblock


\bibitem[Mart\'{\i}nez-Fern\'{a}ndez et~al\mbox{.}(2022)]%
        {10.1145/3487043}
\bibfield{author}{\bibinfo{person}{Silverio Mart\'{\i}nez-Fern\'{a}ndez}, \bibinfo{person}{Justus Bogner}, \bibinfo{person}{Xavier Franch}, \bibinfo{person}{Marc Oriol}, \bibinfo{person}{Julien Siebert}, \bibinfo{person}{Adam Trendowicz}, \bibinfo{person}{Anna~Maria Vollmer}, {and} \bibinfo{person}{Stefan Wagner}.} \bibinfo{year}{2022}\natexlab{}.
\newblock \showarticletitle{Software Engineering for AI-Based Systems: A Survey}.
\newblock \bibinfo{journal}{\emph{ACM Trans. Softw. Eng. Methodol.}} \bibinfo{volume}{31}, \bibinfo{number}{2}, Article \bibinfo{articleno}{37e} (\bibinfo{date}{apr} \bibinfo{year}{2022}), \bibinfo{numpages}{59}~pages.
\newblock
\showISSN{1049-331X}
\href{https://doi.org/10.1145/3487043}{doi:\nolinkurl{10.1145/3487043}}


\bibitem[Nazim et~al\mbox{.}(2020)]%
        {Nazim2020Generating}
\bibfield{author}{\bibinfo{person}{Mohd. Nazim}, \bibinfo{person}{Chaudhary~Wali Mohammad}, {and} \bibinfo{person}{Mohd. Sadiq}.} \bibinfo{year}{2020}\natexlab{}.
\newblock \showarticletitle{Generating {Datasets} for {Software} {Requirements} {Prioritization} {Research}}. In \bibinfo{booktitle}{\emph{2020 {IEEE} {International} {Conference} on {Computing}, {Power} and {Communication} {Technologies} ({GUCON})}}. IEEE, \bibinfo{pages}{344--349}.
\newblock
\href{https://doi.org/10.1109/gucon48875.2020.9231062}{doi:\nolinkurl{10.1109/gucon48875.2020.9231062}}


\bibitem[OpenAI.(2024)]%
        {openai2024gpt4technicalreport}
\bibfield{author}{\bibinfo{person}{OpenAI.}} \bibinfo{year}{2024}\natexlab{}.
\newblock \bibinfo{title}{GPT-4 Technical Report}.
\newblock
\showeprint[arxiv]{2303.08774}~[cs.CL]
\urldef\tempurl%
\url{https://arxiv.org/abs/2303.08774}
\showURL{%
\tempurl}


\bibitem[Ozkaya(2023)]%
        {10109345}
\bibfield{author}{\bibinfo{person}{Ipek Ozkaya}.} \bibinfo{year}{2023}\natexlab{}.
\newblock \showarticletitle{Application of Large Language Models to Software Engineering Tasks: Opportunities, Risks, and Implications}.
\newblock \bibinfo{journal}{\emph{IEEE Software}} \bibinfo{volume}{40}, \bibinfo{number}{3} (\bibinfo{year}{2023}), \bibinfo{pages}{4--8}.
\newblock
\href{https://doi.org/10.1109/MS.2023.3248401}{doi:\nolinkurl{10.1109/MS.2023.3248401}}


\bibitem[Park et~al\mbox{.}(2023)]%
        {s23104949}
\bibfield{author}{\bibinfo{person}{Catherine Park}, \bibinfo{person}{Mohammad~Dehghan Rouzi}, \bibinfo{person}{Md~Moin~Uddin Atique}, \bibinfo{person}{M.~G. Finco}, \bibinfo{person}{Ram~Kinker Mishra}, \bibinfo{person}{Griselda Barba-Villalobos}, \bibinfo{person}{Emily Crossman}, \bibinfo{person}{Chima Amushie}, \bibinfo{person}{Jacqueline Nguyen}, \bibinfo{person}{Chadi Calarge}, {and} \bibinfo{person}{Bijan Najafi}.} \bibinfo{year}{2023}\natexlab{}.
\newblock \showarticletitle{Machine Learning-Based Aggression Detection in Children with ADHD Using Sensor-Based Physical Activity Monitoring}.
\newblock \bibinfo{journal}{\emph{Sensors}} \bibinfo{volume}{23}, \bibinfo{number}{10} (\bibinfo{year}{2023}).
\newblock
\showISSN{1424-8220}
\href{https://doi.org/10.3390/s23104949}{doi:\nolinkurl{10.3390/s23104949}}


\bibitem[Rahman et~al\mbox{.}(2024)]%
        {10795004}
\bibfield{author}{\bibinfo{person}{Tajmilur Rahman}, \bibinfo{person}{Yuecai Zhu}, \bibinfo{person}{Lamyea Maha}, \bibinfo{person}{Chanchal Roy}, \bibinfo{person}{Banani Roy}, {and} \bibinfo{person}{Kevin Schneider}.} \bibinfo{year}{2024}\natexlab{}.
\newblock \showarticletitle{Take Loads Off Your Developers: Automated User Story Generation using Large Language Model}. In \bibinfo{booktitle}{\emph{2024 IEEE International Conference on Software Maintenance and Evolution (ICSME)}}. \bibinfo{pages}{791--801}.
\newblock
\href{https://doi.org/10.1109/ICSME58944.2024.00082}{doi:\nolinkurl{10.1109/ICSME58944.2024.00082}}


\bibitem[Rodeghero et~al\mbox{.}(2017)]%
        {7985649}
\bibfield{author}{\bibinfo{person}{Paige Rodeghero}, \bibinfo{person}{Siyuan Jiang}, \bibinfo{person}{Ameer Armaly}, {and} \bibinfo{person}{Collin McMillan}.} \bibinfo{year}{2017}\natexlab{}.
\newblock \showarticletitle{Detecting User Story Information in Developer-Client Conversations to Generate Extractive Summaries}. In \bibinfo{booktitle}{\emph{2017 IEEE/ACM 39th International Conference on Software Engineering (ICSE)}}. \bibinfo{pages}{49--59}.
\newblock
\href{https://doi.org/10.1109/ICSE.2017.13}{doi:\nolinkurl{10.1109/ICSE.2017.13}}


\bibitem[Sahu et~al\mbox{.}(2024)]%
        {Sahu2024ReqNet}
\bibfield{author}{\bibinfo{person}{Chandan~Kumar Sahu}, \bibinfo{person}{Rahul Rai}, \bibinfo{person}{Margaret Wiecek}, {and} \bibinfo{person}{David Gorsich}.} \bibinfo{year}{2024}\natexlab{}.
\newblock \showarticletitle{ReqNet and {ReqSim}: A {Network} and {Semantic} {Similarity} {Dataset} of {Requirements} from the {Tree} {Structure} of {System} {Requirement} {Specifications}}.
\newblock \bibinfo{journal}{\emph{Journal of Computing and Information Science in Engineering}} (\bibinfo{date}{jun 24} \bibinfo{year}{2024}), \bibinfo{pages}{1--15}.
\newblock
\showISSN{1530-9827}
\href{https://doi.org/10.1115/1.4065786}{doi:\nolinkurl{10.1115/1.4065786}}


\bibitem[Sharma et~al\mbox{.}(2024)]%
        {10724709}
\bibfield{author}{\bibinfo{person}{Amol Sharma}, \bibinfo{person}{Amrita Chaturvedi}, {and} \bibinfo{person}{Anil~Kumar Tripathi}.} \bibinfo{year}{2024}\natexlab{}.
\newblock \showarticletitle{From Problem Descriptions to User Stories: Utilizing Large Language Models through Prompt Chaining}. In \bibinfo{booktitle}{\emph{2024 15th International Conference on Computing Communication and Networking Technologies (ICCCNT)}}. \bibinfo{pages}{1--6}.
\newblock
\href{https://doi.org/10.1109/ICCCNT61001.2024.10724709}{doi:\nolinkurl{10.1109/ICCCNT61001.2024.10724709}}


\bibitem[Team.(2024)]%
        {geminiteam2024geminifamilyhighlycapable}
\bibfield{author}{\bibinfo{person}{Gemini Team.}} \bibinfo{year}{2024}\natexlab{}.
\newblock \bibinfo{title}{Gemini: A Family of Highly Capable Multimodal Models}.
\newblock
\showeprint[arxiv]{2312.11805}~[cs.CL]
\urldef\tempurl%
\url{https://arxiv.org/abs/2312.11805}
\showURL{%
\tempurl}


\bibitem[Touvron et~al\mbox{.}(2023)]%
        {touvron2023llamaopenefficientfoundation}
\bibfield{author}{\bibinfo{person}{Hugo Touvron}, \bibinfo{person}{Thibaut Lavril}, \bibinfo{person}{Gautier Izacard}, \bibinfo{person}{Xavier Martinet}, \bibinfo{person}{Marie-Anne Lachaux}, \bibinfo{person}{Timothée Lacroix}, \bibinfo{person}{Baptiste Rozière}, \bibinfo{person}{Naman Goyal}, \bibinfo{person}{Eric Hambro}, \bibinfo{person}{Faisal Azhar}, \bibinfo{person}{Aurelien Rodriguez}, \bibinfo{person}{Armand Joulin}, \bibinfo{person}{Edouard Grave}, {and} \bibinfo{person}{Guillaume Lample}.} \bibinfo{year}{2023}\natexlab{}.
\newblock \bibinfo{title}{LLaMA: Open and Efficient Foundation Language Models}.
\newblock
\showeprint[arxiv]{2302.13971}~[cs.CL]
\urldef\tempurl%
\url{https://arxiv.org/abs/2302.13971}
\showURL{%
\tempurl}


\bibitem[Vakkuri et~al\mbox{.}(2020)]%
        {10.1109/MS.2020.2985621}
\bibfield{author}{\bibinfo{person}{Ville Vakkuri}, \bibinfo{person}{Kai-Kristian Kemell}, \bibinfo{person}{Joni Kultanen}, {and} \bibinfo{person}{Pekka Abrahamsson}.} \bibinfo{year}{2020}\natexlab{}.
\newblock \showarticletitle{The Current State of Industrial Practice in Artificial Intelligence Ethics}.
\newblock \bibinfo{journal}{\emph{IEEE Softw.}} \bibinfo{volume}{37}, \bibinfo{number}{4} (\bibinfo{date}{jul} \bibinfo{year}{2020}), \bibinfo{pages}{50–57}.
\newblock
\showISSN{0740-7459}
\href{https://doi.org/10.1109/MS.2020.2985621}{doi:\nolinkurl{10.1109/MS.2020.2985621}}


\end{thebibliography}










\appendix
\section*{Appendix}
\section{Domain frequency} 
\label{appen1}
The most frequently occurring domain in the UStAI is Health, appearing 7 times, followed closely by Information Systems with 6 occurrences and Autonomous Vehicles with 4 occurrences. Domains such as Law and Marketing appeared twice each. The remaining domains—including Social Media, Security, Biochemistry, Transportation, Economic, Pattern Recognition, Finance, Social Work, Library and Information Sciences, Movies, Sociology, Demography, Linguistics, Computer Networks, Signal Processing, Sports, Management Information Systems, Music, Social Network, Food, and Urban Studies—each occurred once.

\end{document}